\documentclass[preprint,12pts,
showpacs, 
eqsecnum
]{revtex4}

\newcommand{\be}{\begin{equation}}
\newcommand{\ee}{\end{equation}} 
\newcommand{\bea}{\begin{eqnarray}} 
\usepackage{graphicx}
\newcommand{\eea}{\end{eqnarray}}
\usepackage{epsfig}
\usepackage{bm}

\begin{document}


\title{Vorticity Statistics and the Time Scales of Turbulent Strain}
\author{L. Moriconi}
\affiliation{Instituto de F\'\i sica, Universidade Federal do Rio de Janeiro, \\
C.P. 68528, CEP: 21945-970, Rio de Janeiro, RJ, Brazil}
\author{R.M. Pereira}
\affiliation{Divis\~{a}ão de Metrologia em Din\^{a}mica de Fluidos, Instituto Nacional de Metrologia, 
Normaliza\c{c}\~{a}o e Qualidade Industrial, Av. Nossa Senhora das Gra\c{c}as 50, Duque de Caxias, 
25250-020, Rio de Janeiro, Brazil}

\begin{abstract}
Time scales of turbulent strain activity, denoted as the strain persistence times of first 
and second order, are obtained from time-dependent expectation values and correlation 
functions of lagrangian rate-of-strain eigenvalues taken in particularly defined statistical 
ensembles. Taking into account direct numerical simulation data, our approach relies on heuristic 
closure hypotheses which allow us to establish a connection between the statistics of vorticity 
and strain. It turns out that softly divergent prefactors correct the usual ``$1/s$" strain 
time-scale estimate of standard turbulence phenomenology, in a way which is consistent with the 
phenomenon of vorticity intermittency. 

\end{abstract}

\maketitle

\section{Introduction}

It is a point of reasonable consensus that further progress in the statistical theory of turbulence has been hampered in great part due to the fact that one of its phenomenological pillars -- the Kolmogorov-Richardson cascade -- is actually a longstanding open issue. The usual assumption of eddy stretching as the essential mechanism for the local flow of turbulent kinetic energy towards smaller scales has been challenged by the visualization of multiscale vortical structures in real and numerical experiments \cite{she_etal, douady_etal, bonn_etal, vin_men, ishihara_etal}, and the related discovery of geometrical statistics phenomena \cite{ashurst_etal,tsinober}. One may expect that significative advances in the derivation of the statistical properties of turbulence will follow from a deeper understanding of flow instabilities and their role in the production of coherent structures, within more elaborate discussions of the coupled dynamics of vorticity and the rate-of-strain tensor.

A fundamental problem in this context is to determine for how long a given fluid element is, in its lagrangian evolution, coherently compressed or stretched by the underlying strain field. According to common wisdom \cite{tl}, if $s$ is some measure of the strain strength, such a ``strain persistence time" can be estimated as $T(s) \sim 1/s$.  However, this expression for $T(s)$ is in fact problematic, since the constancy of $sT(s)$ suggests that weak large scale and strong small scale rate-of-strain fluctuations would, respectively, (i) break statistical isotropy at small scales and (ii) have no role in the production of coherent structures, as vortex tubes. Both of these implications are at variance with experimental and numerical observations \cite{comment1}.

Having in mind the above difficulties and relying more on heuristic arguments than on mathematically rigorous grounds, our aim in this work is to suggest that instead of a single time scale $T(s)$, the strain activity can be naturally associated to two distinct time scales, which will be denoted as the strain persistence times of first and second order. It turns out that these time scales contain divergent prefactors which multiply the usual $1/s$ estimate of standard phenomenology, a fact that one may conjecture to be related to the existence of strong vorticity fields and the phenomenon of turbulent intermittency.

This paper is organized as follows. In the next section we address formal definitions of the strain persistence times and discuss, by means of a straightforward closure scheme motivated in great part by the analysis of direct numerical simulation (DNS) data, their relation to single-point vorticity statistics. In Sec. III, we verify, in the DNS context, that our analytical framework, devised to hold in principle in the small strain domain, incidentally holds for the whole range of strain strengths. In Sec. IV, we comment on our findings and point out directions of further research.

\section{Strain Persistence Times}

Let $s_{ij} = (\partial_i v_j + \partial_j v_i)/2$ be the $(i,j)$-component of the lagrangian rate-of-strain tensor. Recalling that $s_{ij}$ is traceless due to incompressibility, call the only positive or the only negative eigenvalue of $s_{ij}(t)$ by $\bar s(t)$, a piecewise continuous function of time, as indicated in Fig. 1a. 

Independent turbulent flow realizations of $\bar s(t)$ generated, for instance, from some set of random initial conditions at $t \rightarrow - \infty$ constitute a large functional space $S$. Take the ensemble $\bar \Lambda_s \subset S$ of all the profiles $\bar s(t)$ which have $\bar s(0)=s$ for an arbitrarily prescribed eigenvalue $s$. Alternatively, we define the related ensemble $\Lambda_s$ of compactly supported functions $s(t)$ which are identified to $\bar s(t) \in \bar \Lambda_s$ in the largest neighborhood of $t=0$ where $\bar s(t)$ is continuous. The functions $s(t)$ vanish out of these neighborhoods. See the sketches in Fig. 1b.

In a more formal way $\Lambda_s$ is given as the ensemble of functions $s(t)$ obtained from the one-to-one mapping
\bea
&&\bar \Lambda_s \mapsto \Lambda_s \nonumber \\
&&\bar s(t) \rightarrow s(t) \ , \ \label{lambdamap1}
\eea
given, for positive $t$, by
\be
s(t) = \left\{ \begin{array}{rcl}
\bar s(t) & \mbox{if}
& \forall t' \in [0,t]  \ , \ \bar s(t')/s > 0  \ , \  \\
0 & \mbox{if}
& \exists t' \in [0,t] \mid \bar s(t')/s < 0  \ , \ \label{lambdamap2}
\end{array} \right. \label{sigma-omega1}
\ee
while for negative $t$, the time interval $[0,t]$ is replaced, in (\ref{sigma-omega1}),
by $[t,0]$. Notwithstanding the mathematically rigorous language used in (\ref{lambdamap1})
and (\ref{lambdamap2}), it is important to keep in mind the essential heuristic-phenomenological
flavor of the present work.

The rationale for the introduction of the ensemble $\Lambda_s$ is that their elements, i.e.,
the time-dependent strain eigenvalues $s(t)$, have all the same postulated ``strain strength" $s \equiv s(0)$
and well-defined lifetimes, once they are compactly supported functions. Our task, therefore, is to 
investigate their characteristic time scales and to understand how they depend on $s$.

\begin{center}
\begin{figure}[tbph]
\vspace{0.0cm}
\includegraphics[width=10.6cm, height=14.8cm]{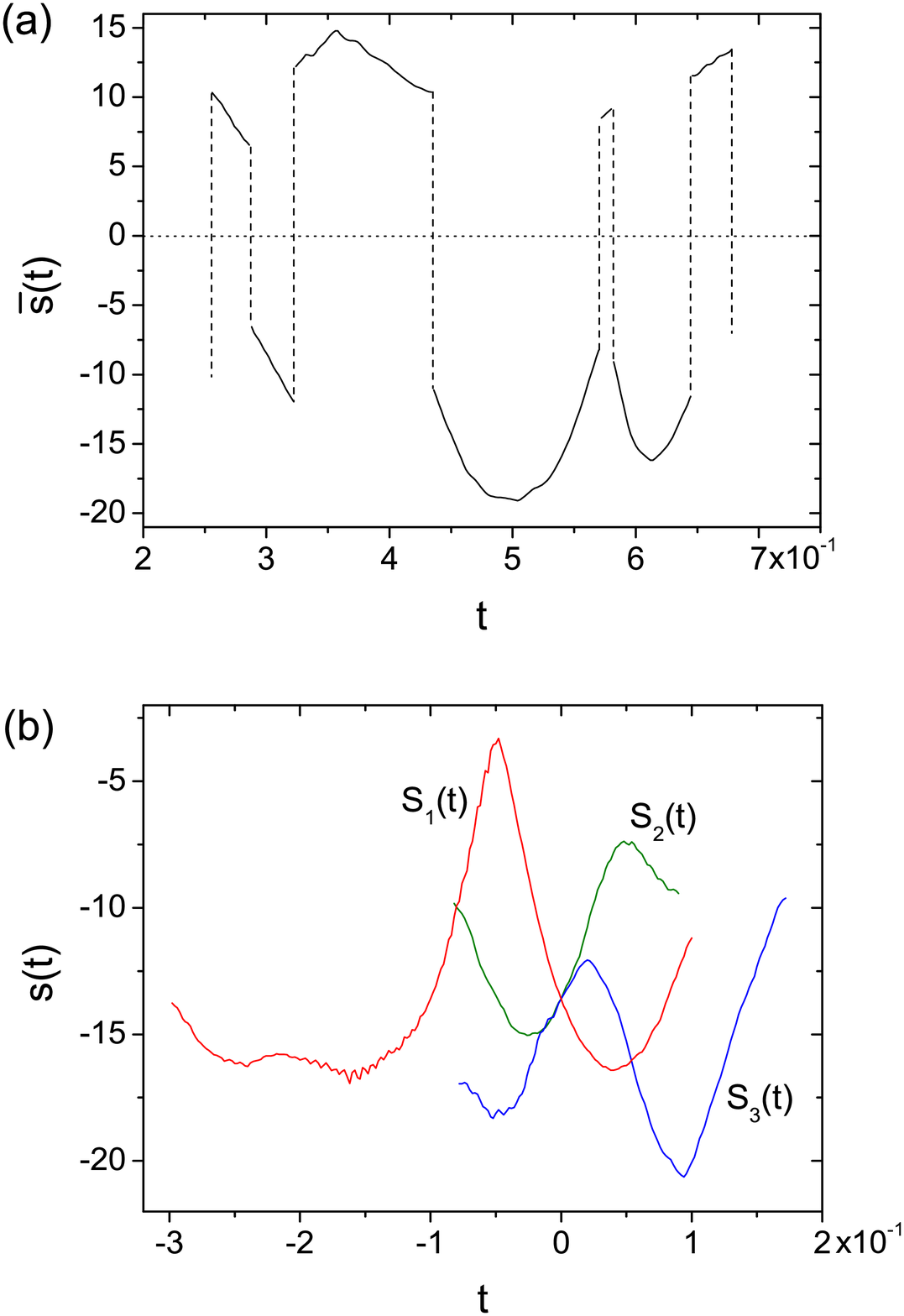}
\label{lambda_s}
\caption{Color online. Figs. (a) and (b) are produced from direct numerical simulation (DNS) data which is addressed in Sec. III. 
In (a) we show a typical profile of the piecewise continuous function $\bar s(t)$. In (b) the compactly supported 
configurations $s_1(t)$ (red), $s_2(t)$ (green) and $s_3(t)$ (blue) belong all to the ensemble $\Lambda_s$, with $s=-13.6 \pm 0.1$, 
since $s_1(0)=s_2(0)=s_3(0)=s$. The total time span of the DNS is $\Delta t = 2048 \times 10^{-3}$ in arbitrary time 
units.}
\vspace{0.0cm}
\end{figure}
\end{center}
\vspace{-1.0cm} 

We emphasize that the just given definition of the ensemble $\Lambda_s$ is a strong idealization of dubious utility if the interest is to compute statistical averages of fluid dynamic observables out of experimental 
or numerical data. The essential difficulty here is that one should work, in principle, with a large functional space of turbulent flow realizations. However, by evoking ergodicity \cite{gal_tsi}, a statistically 
equivalent ensemble $\Lambda_s$ can be introduced for practical purposes as it follows:
\vspace{0.2cm}

(i) Taking a single three-dimensional turbulent flow realization (e.g., the one obtained in a direct numerical simulation) 
recorded in a time interval of length $T$, pick up a number $N$ of (hopefully weakly correlated) lagrangian trajectories; 
\vspace{0.2cm}

(ii) An alternative set of time parametrizations of the lagrangian trajectories is implemented by conventionally setting time $t=0$ at $M$ equally spaced time instants along the dynamical evolution;
\vspace{0.2cm}

(iii) Define the ensemble $\bar \Lambda_s$ of $N \times M$ strain eigenvalue profiles $\bar s(t)$ obtained from the lagrangian 
trajectories introduced in (i) and time-parametrized according to (ii);
\vspace{0.2cm}

(iv) Working with some assigned uncertainty $\delta s$ in the definition of $s$, introduce the ensemble $\Lambda_{s}$ of 
functions $s(t)$ derived from the profiles defined in (iii), as prescribed in (\ref{lambdamap1}) and (\ref{lambdamap2}).
\vspace{0.2cm}

Of particular importance in our considerations are the following time integrations over first and second order 
expectation values taken in $\Lambda_s$,
\bea
&&I_1(s) \equiv \int_{-\infty}^0 dt \langle s(t) \rangle_{\Lambda_s} \ , \ \label{I1} \\
&&I_2(s) \equiv \int_{-\infty}^0 dt \int_{-\infty}^0 dt' \langle \delta s(t) \delta s(t') \rangle_{\Lambda_s} \label{I2} \ , \
\eea
where 
\bea
&&\delta s(t) \equiv s(t) - \langle s(t) \rangle_{\Lambda_s} \ . \
\eea
Considering, for convenience, positive and negative $s$ as separate cases, strain persistence times of first and 
second order, $T^\pm_1(s)$ and $T^\pm_2(s)$, respectively, can be defined from (\ref{I1}) 
and (\ref{I2}), as
\be
I_1(s) = \left\{ \begin{array}{rcl}
sT^+_1(s) & \mbox{for } s>0   \ , \ \\
sT^-_1(s) & \mbox{for } s<0   \ , \
\end{array} \right. \label{eq-s1}
\ee
and
\be
I_2(s) = \left\{ \begin{array}{rcl}
 \mbox{[}sT^+_2(s)\mbox{]}^2 & \mbox{for } s>0  \ , \  \\
 \mbox{[}sT^-_2(s)\mbox{]}^2 & \mbox{for } s<0   \ . \
\end{array} \right. \label{eq-s2}
\ee 

The central idea underlying our discussion is that analytical expressions for $T^\pm_1(s)$ and $T^\pm_2(s)$ can be derived from a statistical treatment of vorticity fluctuations, which are governed by the 
lagrangian evolution equation \cite{pope},
\be
\frac{d \omega_i}{dt} = s_{ij}(t) \omega_j  + \epsilon_{ijk} \partial_j f_k + \nu \partial^2 \omega_i \ , \ \label{ws-lang}
\ee 
where $f_k$ and $\nu$ denote, respectively, the density of external force and the kinematic viscosity.
Let $\hat n(t)$ be the unit vector defined along the principal direction associated to the only positive or only negative eigenvalue $s(t)$ 
of the lagrangian rate-of-strain tensor (the twofold orientation ambiguity of $\hat n(t)$ is arbitrarily resolved). One gets, after some simple 
algebra on Eq. (\ref{ws-lang}),
\be
\frac{d \omega^2 }{dt} = 2 s(t) \omega^2 + 2 \omega \vec \omega \cdot \frac{d \hat n}{dt} + 
\epsilon_{ijk} \omega_i \partial_j f_k
+ \nu \omega_i \partial^2 \omega_i \ , \ \label{w2-lang}
\ee
where $\omega(t) \equiv \vec \omega(t) \cdot \hat n(t)$. It is interesting to take the expectation value of Eq. (\ref{w2-lang}) conditioned to a given time-dependent profile $s(t) \in \Lambda_s$. 
In other words, we change our focus to the alternative evolution equation
\be
\frac{d \langle \omega^2 \rangle_{[s]}}{dt} = 2 s(t) \langle \omega^2 \rangle_{[s]} + 
2 \langle \omega \vec \omega \cdot \frac{d \hat n}{dt}  \rangle_{[s]} +
\epsilon_{ijk} \langle \omega_i \partial_j f_k \rangle_{[s]}
+ \nu \langle \omega_i \partial^2 \omega_i \rangle_{[s]}  \ , \ \label{w2s-lang}
\ee
where $\langle (...) \rangle_{[s]}$ is the self-evident notation for the procedure of conditional averaging. It is clear that 
Eq. (\ref{w2s-lang}) is not closed. The second term on its right-hand-side, for instance, is related in an intrincate way, through $d \hat n / dt$, 
to the velocity gradient tensor and the pressure hessian. Neglecting small alignment effects \cite{ashurst_etal}, we take, as a first approximation, that $\vec \omega(t)$ and $\hat n(t)$ are 
completely uncorrelated in the ensemble of flow realizations which share the same arbitrary profile $s(t)$. Resorting furthermore to isotropy, it follows, thus, that
\be
\langle \omega \vec \omega \cdot \frac{d \hat n}{dt}  \rangle_{[s]}
= \langle \omega_i \omega_j \rangle_{[s]} \langle \hat n \cdot \hat e_i \frac{d}{dt} (\hat n \cdot \hat e_j)  \rangle_{[s]} 
= \frac{1}{3} \langle \vec \omega^2 \rangle_{[s]} \langle \hat n \cdot \frac{d \hat n}{dt} \rangle_{[s]} = 0 \ . \
\label{wwnn}
\ee
The last two terms in Eq. (\ref{w2s-lang}) are just the rates of enstrophy injection and dissipation, so that their combined contribution is assumed to vanish. 
Taking Eq. (\ref{wwnn}) into account, we are led, after a straightforward integration of Eq. (\ref{w2s-lang}), to
\be
\langle \omega^2(t) \rangle_{[s]} = \langle \omega^2(T) \rangle_{[s]} \exp \left [ 2\int_T^t dt' s(t') \right ]  \ . \ \label{w2s}
\ee
Eq. (\ref{w2s}) is now averaged over the configurations $s(t) \in \Lambda_s$. This can be rethorically expressed in path-integral
language \cite{feynman-hibbs} as
\bea
\langle \omega^2(0) \rangle_{\Lambda_s} &=& \int_{\Lambda_s} D[s(t)] \rho[s(t)]
\langle \omega^2(T) \rangle_{[s]} \exp \left [ 2\int_T^0 dt' s(t') \right ]  \nonumber \\
&=& \int_{\Lambda_s} D[s(t)] \rho[s(t)]
 \exp \left [ 2\int_T^0 dt' s(t') + \ln \langle \omega^2(T) \rangle_{[s]} \right ]  \ , \ \label{path_int}
\eea
where $D[s(t)] \rho[s(t)]$ is the probability measure defined on $\Lambda_s$. The time $T<0$ in the above equation
is actually an arbitrary parameter, which we take to be the largest time instant where $s(t)=0$ (that is, $T$ is a 
negative-valued functional of $s(t)$ in the ensemble $\Lambda_s$).
Consider, now, the random variables
\bea
\xi &\equiv&  \int_T^0 dt' s(t') \ , \ \\ 
\tilde \xi &\equiv& \int_T^0 dt' s(t') + \frac{1}{2} \ln \langle \omega^2(T) \rangle_{[s]} \ . \ \label{xi-tilde-xi}
\eea
It is clear, from (\ref{w2s}) and (\ref{xi-tilde-xi}), that $\langle \omega^2(0) \rangle_{[s]} = \exp[2 \tilde \xi ]$. Despite the fact that the statistical 
properties of $\xi$ and $\tilde \xi$ are analytically unknown, we claim that their first and second order moments are related 
in a simple linear way, as
\bea
\langle \tilde \xi \rangle &=& \alpha \langle \xi \rangle + c\ , \ \label{mean_xi} \\ 
\langle (\tilde \xi - \langle \tilde \xi \rangle)^2 \rangle &=& \beta \langle (\xi - \langle \xi \rangle)^2 \rangle \ . \ \label{var_xi}
\eea
Above, $\alpha$, $\beta$ and $c$ are constant parameters, which we need to determine. It is important to stress that Eqs. (\ref{mean_xi}) 
and (\ref{var_xi}) are fundamental closure hypotheses which lead, as we will see, to a reasonable account of statistical quantities 
evaluated from direct numerical simulation data.

Taking into account (\ref{mean_xi}) and (\ref{var_xi}), the cumulant expansion method \cite{crews_etal} can be straightforwardly applied 
to Eq. (\ref{path_int}) to yield, up to second order in the strain fluctuations,
\be
\langle \omega^2(0) \rangle_{\Lambda_s} = \exp \left [2c +  2 \alpha \int_{-\infty}^0 dt \langle s(t) \rangle_{\Lambda_s}  + 
2 \beta \int_{-\infty}^0 dt \int_{-\infty}^0 dt' \langle \delta s(t) \delta s(t') \rangle_{\Lambda_s} \right ]  \label{s_dsds}
\ , \
\label{c-expansion}
\ee
Using Eqs. (\ref{I1}) to (\ref{eq-s2}), the right hand side of Eq. (\ref{c-expansion}) can be rewritten as
\be 
\langle \omega^2(0) \rangle_{\Lambda_s} =
\left\{ \begin{array}{rcl}
 \exp \left \{ 2c+ 2 \alpha s T^+_1(s) + 2 \beta [s T^+_2(s)]^2 \right \}  & \mbox{for}
& s>0 \ , \ \\
 \exp \left \{ 2c+ 2 \alpha s T^-_1(s) + 2 \beta [s T^-_2(s)]^2 \right \}& \mbox{for}
& s<0 \ . \ 
\end{array} \right.
\label{omega-TT}
\ee

It is also convenient to express, up to first order in a power series of $s$, the standard deviation of $\omega(0)$ in the 
ensemble $\Lambda_s$ as
\be
\sigma_\omega(s) \equiv \sqrt{\langle \omega^2(0) \rangle_{\Lambda_s}} = \left\{ \begin{array}{rcl}
a(1 + b_+ s) & \mbox{for}
& s>0 \ , \  \\
a(1 - b_- s) & \mbox{for}
& s<0 \ , \
\end{array} \right.  \label{sigma-omega}
\ee
where $a>0$, $b_+$ and $b_-$ are arbitrary coefficients and we have used that $\vec \omega$ and $\hat n$ are independent random variables in $\Lambda_s$, 
so that $\langle \omega(0) \rangle_{\Lambda_s} = 0$. As reported in the next section, the relevance of the above expansion is established in a purely 
empirical way (that is, from the analysis of DNS data) with the surprising result -- still in need of theoretical understanding -- that there are no 
higher order corrections to Eq. (\ref{sigma-omega}) even for reasonably large values of the strain eigenvalue $s$.

As stated (in rephrased form) in the introductory section, we do not expect vanishing strain to have any effect on the statistics of vorticity. 
Requiring, therefore, that 
\be
\lim_{s \rightarrow 0} [s T^\pm_1(s)] = \lim_{s \rightarrow 0} [s T^\pm_2(s)]=0 \ , \
\ee
it follows from Eqs. (\ref{omega-TT}) and (\ref{sigma-omega}) that $\exp(c)=a$. 

Recalling, now, the time-reversal symmetry of the fluid dynamic equations in the absence of forcing and dissipation terms, 
a meaningful approximation for the description of inertial range processes, we assume that $T^+_1(s) \propto  T^-_1(-s)$ and
$T^+_2(s) \propto  T^-_2(-s)$. We point out that this argument is not inconsistent at all with 
the dissipation anomaly postulated by the ``zeroth law" of turbulence, that is, the fact that 
energy dissipation rate per unit volume is finite in the inviscid limit 
$\nu \rightarrow 0$ \cite{pope, frisch, davidson, eyink, kaneda_etal}. The situation here is 
analogous to the issue on the coexistence of the second law of thermodynamics with microscopic 
reversibility in the statistical mechanics context. Introducing a pair of even functions of $s$, 
$F_1(s)$ and $F_2(s)$, and proportionality constants $g$ and $g'$, we may write, thus, without 
loss of further generality, that
\bea
&&T_1^+ (s) = \frac{1}{\alpha s} \ln F_1(s) \ , \ T_1^- (s) = - \frac{g}{\alpha s} \ln F_1(s) \ , \ \label{AT}  \\
&&[T_2^+ (s)]^2 = \frac{1}{\beta s^2} \ln F_2(s) \ , \ [T_2^- (s)]^2 = \frac{g'}{\beta s^2} \ln F_2(s) \label{BT} \ . \
\eea
Note that 
\be
\frac{I_1(-s)}{I_1(s)} = \frac{\int_{-\infty}^0 dt \langle s(t) \rangle_{\Lambda_{-s}} }{\int_{-\infty}^0 dt \langle s(t) \rangle_{\Lambda_s}}=
- \frac{T_1^- (-s)}{T_1^+ (s)} = - g \label{I1I1}
\ee
and 
\be
\frac{I_2(-s)}{I_2(s)} = \frac{\int_{-\infty}^0 dt \int_{-\infty}^0 dt' \langle \delta s(t) \delta s(t') \rangle_{\Lambda_{-s}} }
{\int_{-\infty}^0 dt \int_{-\infty}^0 dt' \langle \delta s(t) \delta s(t') \rangle_{\Lambda_s}}
= \left [ \frac{T_2^- (-s)}{T_2^+ (s)} \right ]^2 = g' \ . \  \label{I2I2}
\ee
Taking Eqs. (\ref{omega-TT}), (\ref{AT}) and (\ref{BT}) and the even parity of $F_1(s)$ and $F_2(s)$ into account, 
it turns out that Eq. (\ref{sigma-omega}) holds if and only if
\bea
&&F_1(s) \cdot F_2(s) = 1 + b_+ |s| \ , \  \label{F1} \\
&&\frac{[F_2(s)]^{g'}}{[F_1(s)]^g} = 1 + b_-|s| \label{F2} \ . \ 
\eea
Substituting the solutions of Eqs. (\ref{F1}) and (\ref{F2}) for $F_1(s)$ and $F_2(s)$ into Eqs. (\ref{AT}) and (\ref{BT}), 
we find
\bea
&&T^+_1(s)= \frac{1}{\alpha (g + g')} T_1(s) \ , \ T^-_1(s)= \frac{g}{\alpha (g + g')} T_1(s) \ , \ \label{TP1} \\
&&T^+_2(s)= \sqrt{\frac{1}{\beta (g + g')}} T_2(s) \ , \ T^-_2(s)= \sqrt{\frac{g'}{\beta (g + g')}} T_2(s) \ , \ \label{TP2}
\eea
where
\bea
&&T_1(s) = \frac{1}{|s|} \ln \left [ \frac{(1+b_+ |s|)^{g'}}{1+b_-|s|} \right ] \ , \ \label{T1} \\
&&T_2(s) = \frac{1}{|s|} \sqrt{\ln \left [ (1+b_+|s|)^g (1+b_-|s|) \right ]} \label{T2} \ . \
\eea
It is interesting to remark that once $T^\pm_1(s)$ is positive definite, it is necessary to have, according to
(\ref{T1}), $g' \geq 1$ and $g'b_+ >b_-$ \cite{comment2}. This is a simple and well-defined prediction from the 
present formalism. 

In the weak strain regime, one expects that the strain persistence times saturate at the large eddy turnover time 
$T_0 \equiv \sqrt[3]{L^2/\epsilon}$, where $L$ and $\epsilon$ are, respectively, the typical large length scale and 
the energy dissipation rate parameters of the turbulent flow. We get, thus, from Eqs. (\ref{TP1}) to (\ref{T2}),
\bea
&&\lim_{s \rightarrow 0} T^\pm_1(s) \propto g'b_+ - b_- \propto T_0 \ , \ \\
&&\lim_{s \rightarrow 0} T^\pm_2(s) \propto gb_+ + b_- \propto T_0 \ . \
\eea
Both parameters $b_+$ and $b_-$ are, therefore, likely to be proportional to the large eddy turnover time $T_0$. 
An interesting problem, not touched here, is to find the Reynolds number dependence, if any, of the dimensionless parameters
$g$, $g'$, $b_+/T_0$, and $b_-/T_0$.  

\section{Analysis of DNS Data}

We have computed statistical averages of fluid dynamic observables with the help of the direct numerical simulation (DNS) database available from the turbulence research group at Johns Hopkins University \cite{jhu}. An homogeneous and isotropic turbulent flow with Taylor-based Reynolds number $R_\lambda \approx 433$ is simulated in a periodic cube of linear dimension $L= 2 \pi$ modeled as a grid of $1024^3$ lattice points. Viscosity and time step parameters are, respectively, $\nu = 1.85 \times 10^{-4}$ and $\Delta t = 2 \times 10^{-4}$ (the complete simulation record corresponds to around one large eddy turnover time, $T_0 =  2^{10} \times 10 \times \Delta t$). Further simulation details can be found in Refs. \cite{jhu1,jhu2}.

Statistical samples were produced in two different ways, according to the particular expectation values we were interested to evaluate:
\vspace{0.2cm}

(i) In order to compute $\sigma_\omega(s) = \sqrt{\langle \omega^2(0) \rangle_{\Lambda_s}}$, vorticity vectors and rate-of-strain tensors were defined from the velocity gradients taken at grid points $2 \pi (i,8j,8k)/1024$ where $0 \leq i < 1024$ and $0 \leq j,k < 128$ are integer numbers, for $10^3$ frames of equally time-spaced flow configurations. Bins of variable sizes were considered for the sets of positive and negative rate-of-strain eigenvalues $s$. While essentially conventional, our particular bin size choice, $\delta s = 0.5$, proved to yield robust results at economical computational costs.
\vspace{0.2cm}

(ii) $I_1(s)$ and $I_2(s)$, as given in Eqs. (\ref{I1}) and (\ref{I2}), were computed from $10^3$ lagrangian trajectories, each one consisting of $2^{10}$ time steps (which are separated in time by $2 \times 10^{-3}$ arbitrary time units). 
The initial points of the lagrangian trajectories are given by $2 \pi (i,j,k)/10$, with $0 \leq i,j,k \leq 10$. The ensemble $\Lambda_s$, 
with uncertainty $\delta s = 0.1$, was generated through the procedure previously discussed in Sec. II.
\vspace{0.2cm}

\begin{center}
\begin{figure}[tbph]
\vspace{-0.5cm}
\hspace{-0.5cm} 
\includegraphics[width=10.14cm, height=6.97cm]{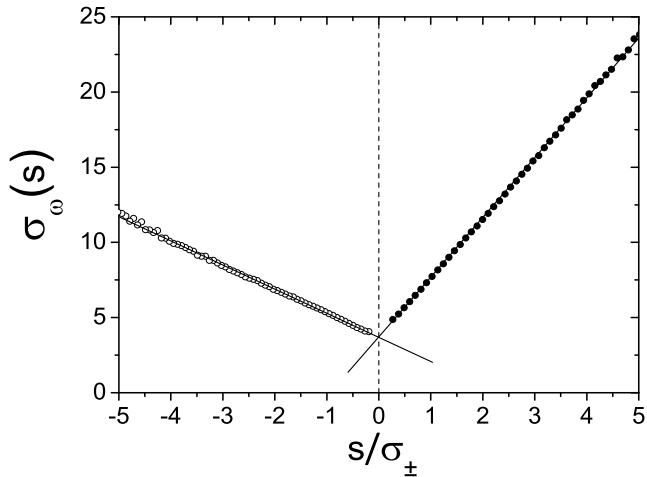}
\label{}
\caption{$\sigma_\omega (s) = \sqrt{\langle \omega^2(0) \rangle_{\Lambda_s}}$, i.e., the standard deviation of the projected lagrangian vorticity $\omega(0)$, is plotted as a function of standardized rate-of-strain eigenvalues $s/\sigma_\pm$. The standard deviations of the positive and negative rate-of-strain eigenvalues $s$ are, respectively, $\sigma_+ = 4.63$ and $\sigma_- = 6.52$.}
\end{figure}
\end{center}
\vspace{-1.0cm}

In both of the above cases (i) and (ii), the eigenvalues and principal directions of the rate-of-strain tensor were computed through an efficient hybrid algorithm which combines direct analytical evaluation and the so-called QL algorithm \cite{diag}.

\begin{center}
\begin{figure}[tbph]
\vspace{-0.0cm}
\hspace{-0.5cm} 
\includegraphics[width=10.2cm, height=14.4cm]{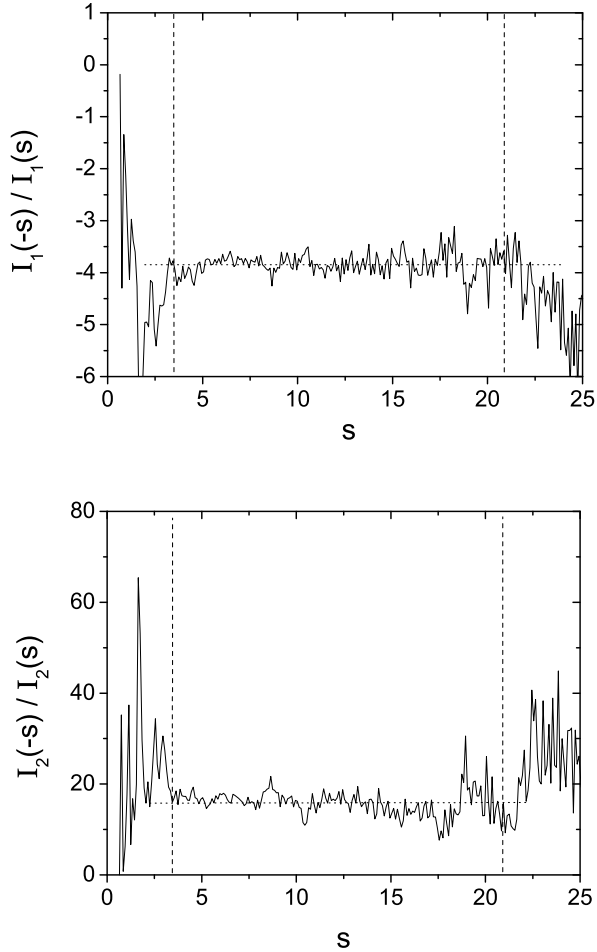}
\label{}
\caption{The horizontal dotted lines indicate the mean values 
$-3.84$ ($\equiv -g$) and $15.81$ ($\equiv g'$) of the ratios $I_1(-s)/I_1(s)$ 
and $I_2(-s)/I_2(s)$, which are, in fact, approximately constant 
in the range $3.4 \leq s \leq 21.0$, the region between vertical 
dashed lines, where the ensembles $\Lambda_s$ are large enough for 
the evaluation of reasonable statistical averages.}
\end{figure}
\end{center}
\vspace{0.0cm}

As it is clear from Fig. 2, the conditional expectation value $\sigma_\omega(s)$ is precisely -- and surprisingly well -- described by Eq. (\ref{sigma-omega}), 
with no additional corrections. At $s=0$ we have $a = \sigma_s(0) = 3.6$. The slope parameters $b_+$ and $b_-$ associated to the right and left branches of 
$\sigma_\omega(s)$, respectively, are $b_+ = a^{-1} d \sigma_\omega(s) / ds |_{s>0} = 0.26$ and $b_- = a^{-1} d \sigma_\omega(s) / ds |_{s<0} = 0.08$.

\begin{center}
\begin{figure}[tbph]
\vspace{-0.0cm}
\hspace{-0.5cm} 
\includegraphics[width=10.2cm, height=14.4cm]{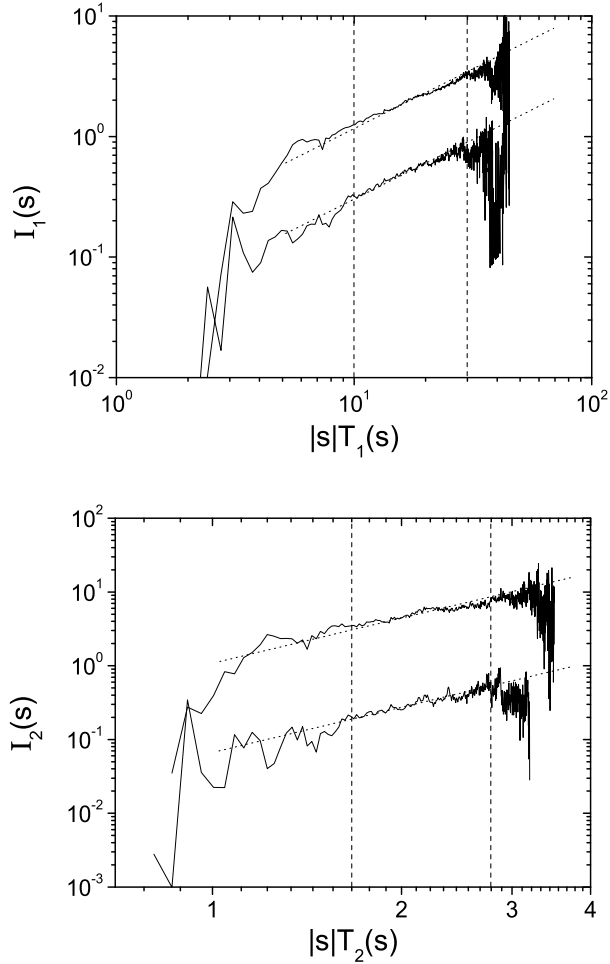}
\label{}
\caption{Plots which provide support for the validity of Eqs. (\ref{I1}) and (\ref{I2}) with the use of 
the strain persistence times $T_1^\pm(s)$ and $T_2^\pm(s)$, as defined from Eqs. (\ref{TP1})-(\ref{T2}). 
The dotted lines in the plots for $I_1(s)$ have unit slope, while the ones in the plots for 
$I_2(s)$ have slope 2. The upper plots in each pair of plots correspond to negative $s$.
The vertical dashed lines indicate the range $3.4 \leq s \leq 21.0$ (see Fig. 3).}
\end{figure}
\end{center}
\vspace{0.0cm}

In Fig. 3, we show that the expectation value ratios $I_1(-s)/I_1(s)$ and $I_2(-s)/I_2(s)$ are approximately constant for an extended range of $s$-values, in agreement with Eqs. (\ref{I1I1}) and (\ref{I2I2}). The measured values of $b_+$, $b_-$, $g = |I_1(-s)/I_1(s)|$ and $g' = I_2(-s)/I_2(s)$ are then substituted in the logarithmic corrections written down in Eqs. (\ref{T1}) and (\ref{T2}), which are strikingly confirmed from the plots shown in Fig. 4.

The absolute ratio $|I_1(-s)/I_1(s)|=T_1(-s)/T_1(s) \simeq 3.84$ is actually expected to be a number larger than unity. This follows from the well-known fact that $s$ is most of the time negative, in other words, $T_1(-s) > T_1(s)$. We have actually verified that the domain of negative $s$ in physical space constitutes around $75\%$ of the total fluid volume, which is equivalent to say that the intermediate rate-of-strain eigenvalue is positively skewed \cite{ashurst_etal, tsinober_etal}.

We note, furthermore, that the estimate $g' \simeq 15.81$ is compatible with the relation $g' b_+ > b_-$, predicted at the end of Sec. III.
\vspace{0.3cm}

{\leftline{\it{Estimation of the $\alpha$ and $\beta$ parameters}}}
\vspace{0.3cm}

The results depicted in Figs. 3 and 4 do not depend on the specific values of the parameters $\alpha$ and $\beta$ introduced in Eqs. (\ref{mean_xi}) and (\ref{var_xi}). 

Using relations (\ref{eq-s1}), (\ref{eq-s2}) , (\ref{TP1})-(\ref{T2}), it follows that at $|s|T_1(s) = 1$,
we have
\be
I_1^+ = \frac{1}{\alpha (g+g')} \ , \ I_1^- = \frac{g}{\alpha (g+g')} \ , \
\ee
while at $|s|T_2(s)=1$,
\be
I_2^+ = \frac{1}{\beta (g+g')} \ , \ I_1^- = \frac{g'}{\beta (g+g')} \ . \
\ee
Therefore, recalling that $g$ and $g'$ have already been determined, the values of $\alpha$ and $\beta$ can be 
straightforwardly computed from the intercepts of the dotted lines in Fig. 4 with the vertical lines $|s|T_1(s)=|s|T_2(s)=1$.
Proceeding in this way, we get $\alpha =1.64$ and $\beta = 0.73$.

Substituting the estimated values of $\alpha$, $\beta$, $g$ and $g'$ in the analytical expressions (\ref{TP1}) and (\ref{TP2}), we 
then perform a consistency check for the definitions of $I_1^\pm(s)$ and $I_2^\pm(s)$, Eqs. (\ref{eq-s1}) and (\ref{eq-s2}). 
The results, shown in Fig. 5, are reasonable accurate, with better perfomance for smaller values of $s$. For larger values $s$,
fluctuations errors come into play spoiling the comparison between analytical expressions and numerical evaluations.

\begin{center}
\begin{figure}[tbph]
\vspace{-0.0cm}
\hspace{-0.5cm} 
\includegraphics[width=10.2cm, height=14.4cm]{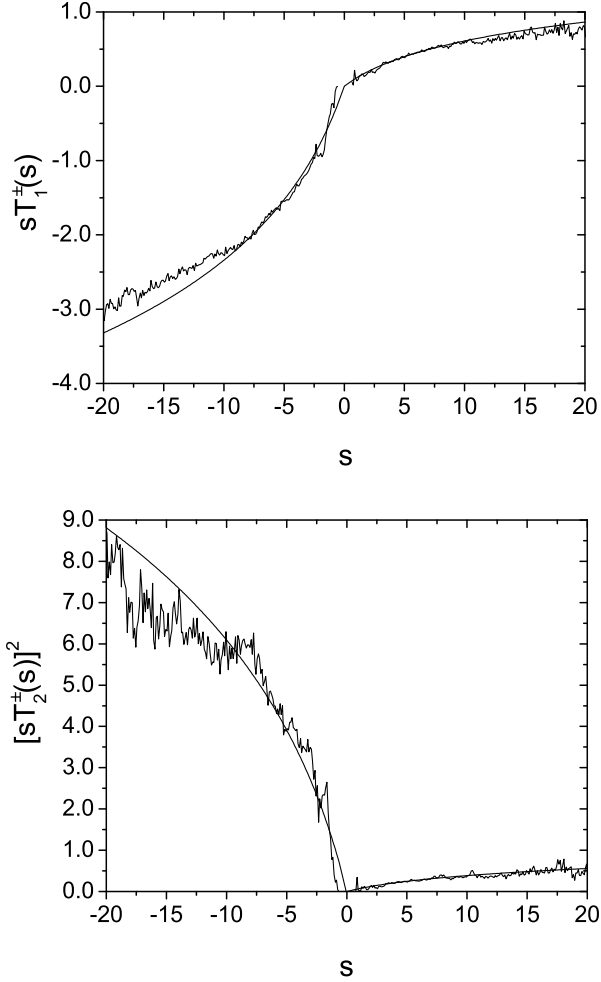}
\label{}
\caption{A test of the relations (\ref{eq-s1}) and (\ref{eq-s2}), where $T_1^{\pm}(s)$ and $T_2^{\pm}(s)$
are given by (\ref{TP1}) and (\ref{TP2}) with $\alpha = 1.64$, $\beta= 0.73$, $g=3.84$, and $g'=15.81$.}
\end{figure}
\end{center}
\vspace{0.0cm}

\begin{center}
\begin{figure}[tbph]
\vspace{-0.0cm}
\hspace{-0.5cm} 
\includegraphics[width=10.2cm, height=6.97cm]{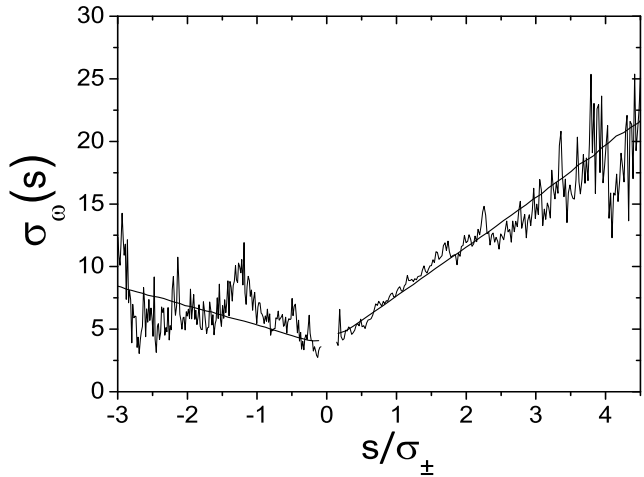}
\label{}
\caption{The numerical evaluation of the square root of the right hand side of (\ref{s_dsds}), which is assumed 
to yield $\sigma_\omega(s)$, is compared to the result previously reported in Fig.2 (straight lines).}
\end{figure}
\end{center}
\vspace{0.0cm}

If, now, the numerically evaluated functions $I_1^\pm(s)$ and $I_2^\pm(s)$ take the place of $sT_1^{\pm}(s)$ and $[sT_2^{\pm}(s)]^2$, respectively,
in Eq. (\ref{s_dsds}), we would expect to recover, from an entirely alternative perspective, the empirical linear profiles depicted in Fig. 2.
In fact, we find, as shown in Fig. 6, suggestive agreement for positive $s$. For negative $s$, such an evaluation of $\sigma_s(\omega)$
is plagued with a stronger numerical uncertainty. This happens ultimately due to the fact that $g$ and $g'$ are both larger than unity. Therefore,
$I_1^-(s)$ and $I_2^-(s)$ have larger error bars than $I_1^+(s)$ and $I_2^+(s)$, respectively, which are exponentially propagated in the
computation of $\sigma_s(\omega)$.  We note, furthermore, that in Fig. 6 the range of $s/\sigma_\pm$ values is a bit smaller than the one
used in Fig. 2. The range of $s/\sigma_\pm$ in Fig. 6 is actually determined by the vertical dashed lines shown (and discussed) in Fig. 3.

\section{Conclusions}

The essential guideline underlying our analysis is that vorticity statistics is the ideal setting for the study of key aspects of the rate-of-strain tensor dynamics. We have devised, from the vorticity field, a suitable conditioned expectation value, $\sigma_\omega(s) \equiv \sqrt{\langle \omega^2(0) \rangle_{\Lambda_s}}$, which is directly related to the time scales of lagrangian strain activity. As a refinement of standard phenomenology, it turns out that two time scales -- the strain persistence times $T_1(s)$ and $T_2(s)$ -- are necessary to accurately reproduce $\sigma_\omega(s)$, as determined from DNS data. The strain persistence times $T_1(s)$ and $T_2(s)$ are introduced as first and second order contributions within a second order cumulant expansion, once closure and working hypotheses have been put forward.

While $|s|T_1(s)$ and $|s|T_2(s)$ vanish by construction at $s =0$, they are both softly divergent at asymptotically large $|s|$, which happens to be a crucial ingredient in the derivation of the linear profiles of $\sigma_\omega(s)$, strikingly indicated in Fig. 2. The divergences of $|s|T_1(s)$ and $|s|T_2(s)$ as $s \rightarrow \pm \infty$ could bring some light on the understanding of the phenomenon of turbulent intermittency, since they suggest that strong -- and hence small scale -- strain fluctuations are likely to have a non-negligible role in the statistical properties of vorticity. It is possible that the second-order truncation in the cumulant expansion (\ref{c-expansion}) is actually a fine approximation to the full non-perturbative result, due mainly to the specific definition of the statistical ensembles $\Lambda_s$, which may provide a partition of the whole functional space into subspaces of gaussian stochastic processes $s(t)$. A point in favor of the second order cumulant expansion is the fact that the ratio between the second order and first order contributions, $\beta [sT_2(s)]^2/ \alpha|s|T_1(s)$, converges to $\beta (g+1)/\alpha (g'-1) \simeq 0.14 < 1$ as $s \rightarrow \pm \infty$.

We highlight that from a purely theoretical perspective, no considerations have been advanced to establish the form of the strain persistence times beyond the first order in $s$. However, we have found that the empirical evaluation of $\sigma_\omega(s)$ does not bring any further non-linear corrections into scene, a fact that seems to be far from trivial (note that statistical isotropy just implies that $\langle \vec \omega^2 \rangle = 2 \langle s_{ij}^2 \rangle$, which looks like a necessary but by no means a sufficient condition for the specific observed profile of $\sigma_\omega(s)$).

There is, of course, a number of assumptions we have made throughout the paper; while they turned out to lead to reasonably good predictions of numerical results, it is clear that further experimental and numerical investigations are in order to fully support them. A natural direction of research is to check to what extent the premises and results proposed here can match the phenomenology implied by promising effective lagrangian simulations of the velocity gradient tensor, as, for instance, the ones carried out within the Recent Fluid Deformation Closure model \cite{chevi1,chevi2}.
\vspace{0.4cm}

{\leftline{\small{\bf{ACKNOWLEDGMENTS}}}}
\vspace{0.4cm}

The authors acknowledge interesting comments of Arkady Tsinober in the initial stage of this work, and
enlightening remarks by Laurent Chevillard and Charles Meneveau. This work has been partially supported 
by CNPq and FAPERJ and was completed in the stimulating scientific environment at NIDF-UFRJ.


\begin{thebibliography}{99}
\bibitem{comment1} To be fair to the usual arguments, the undesirable divergence of $T(s) \sim 1/|s|$ as $s \rightarrow 0$
has not been a source of much concern, since the main focus is, in general, the role of strain in inertial range scales.
\bibitem{comment2} Straightforward proof of the positive-definiteness of $T^\pm_1(s)$: taking a look at Eqs. (\ref{I1}) and (\ref{eq-s1}), $T^\pm_1(s)>0$ follows from the fact that $\langle s(t) \rangle_{\Lambda_s}$ has always the same sign as the rate-of-strain eigenvalue $s(0)=s$.
\end{thebibliography}
\end{document}